\documentclass[sigconf]{acmart}

\PassOptionsToPackage{table,xcdraw}{xcolor} 

\usepackage{hyperref}
\usepackage{booktabs}
\usepackage{multirow}
\usepackage{subcaption}
\usepackage{xcolor,colortbl}
\definecolor{gray}{rgb}{0.1,0.1,0.1}
\usepackage[T1]{fontenc}
\usepackage[english]{babel}
\usepackage{graphicx}
\usepackage[flushleft]{threeparttable}
\usepackage{adjustbox}
\usepackage{subcaption}
\usepackage{longtable}
\usepackage{enumitem}
\usepackage{tabularx}

\AtBeginDocument{%
  \providecommand\BibTeX{{%
    \normalfont B\kern-0.5em{\scshape i\kern-0.25em b}\kern-0.8em\TeX}}}

\copyrightyear{2024}
\acmYear{2024}
\setcopyright{rightsretained}
\acmConference[CSCW Companion '24] {Companion of the 2024 Computer-Supported Cooperative Work and Social Computing}{November 9--13, 2024}{San Jose, Costa Rica.}
\acmBooktitle{Companion of the 2024 Computer-Supported Cooperative Work and Social Computing (CSCW Companion '24), November 9--13, 2024, San Jose, Costa Rica}
\acmISBN{979-8-4007-1114-5/24/11}
\acmDOI{10.1145/3678884.3681833}

\begin{document}
\title[Envisioning New Futures of Positive Social Technology: Beyond Paradigms of Fixing, \\ Protecting, and Preventing]{Envisioning New Futures of Positive Social Technology: Beyond Paradigms of Fixing, Protecting, and Preventing}

\author{JaeWon Kim}
\authornote{Both authors contributed equally to this research.}
\orcid{0000-0003-4302-3221}
\affiliation{%
  \institution{University of Washington}
  \city{Seattle}
  \state{WA}
  \country{USA}}
\email{jaewonk@uw.edu}

\author{Lindsay Popowski}
\authornotemark[1]
\orcid{0000-0002-5649-0286}
\affiliation{%
  \institution{Stanford Univeristy}
  \city{Stanford}
  \state{CA}
  \country{USA}
}
\email{popowski@stanford.edu}

\author{Anna Fang}
\orcid{0000-0003-0055-3011}
\affiliation{%
  \institution{Carnegie Mellon University}
  \city{Pittsburgh}
  \state{PA}
  \country{USA}
}
\email{amfang@andrew.cmu.edu}

\author{Cassidy Pyle}
\orcid{0000-0003-4578-0226}
\affiliation{%
  \institution{University of Michigan}
  \city{Ann Arbor}
  \state{MI}
  \country{USA}
}
\email{cpyle@umich.edu}

\author{Guo Freeman}
\orcid{0000-0001-5107-7794}
\affiliation{%
  \institution{Clemson University}
  \city{Clemson}
  \state{SC}
  \country{USA}
}
\email{guof@clemson.edu}

\author{Ryan M. Kelly}
\orcid{0000-0002-8773-6656}
\affiliation{%
  \institution{RMIT University}
  \city{Melbourne}
  \state{VIC}
  \country{Australia}
}
\email{ryan.kelly@rmit.edu.au}

\author{Angela Y. Lee}
\orcid{0000-0002-9527-5730}
\affiliation{%
  \institution{Stanford University}
  \city{Stanford}
  \state{CA}
  \country{USA}
}
\email{angela8@stanford.edu}

\author{Fannie Liu}
\orcid{0000-0002-5656-3406}
\affiliation{%
  \institution{JPMorgan Chase \& Co.}
  \city{New York}
  \state{NY}
  \country{USA}
}
\email{fannie.liu@jpmchase.com}

\author{Angela D. R. Smith}
\orcid{0000-0001-5546-5452}
\affiliation{%
  \institution{The University of Texas at Austin}
  \city{Austin}
  \state{TX}
  \country{USA}
}
\email{adrsmith@utexas.edu}

\author{Alexandra To}
\orcid{0000-0001-7298-7607}
\affiliation{%
  \institution{Northeastern University}
  \city{Boston}
  \state{MA}
  \country{USA}
}
\email{a.to@northeastern.edu}

\author{Amy X. Zhang}
\orcid{0000-0001-9462-9835}
\affiliation{%
  \institution{University of Washington}
  \city{Seattle}
  \state{WA}
  \country{USA}
}
\email{axz@cs.uw.edu}

\renewcommand{\shortauthors}{JaeWon Kim et al.}

\begin{abstract}
    Social technology research today largely focuses on mitigating the negative impacts of technology and, therefore, often misses the potential of technology to enhance human connections and well-being. However, we see a potential to shift towards a holistic view of social technology's impact on human flourishing. We introduce \textbf{\textit{Positive Social Technology (Positech)}}, a framework that shifts emphasis toward leveraging social technologies to support and augment human flourishing. This workshop is organized around three themes relevant to Positech: 1) ``Exploring Relevant and Adjacent Research'' to define and widen the Positech scope with insights from related fields, 2) ``Projecting the Landscape of Positech'' for participants to outline the domain's key aspects and 3) ``Envisioning the Future of Positech,'' anchored around strategic planning towards a sustainable research community. Ultimately, this workshop will serve as a platform to shift the narrative of social technology research towards a more positive, human-centric approach. It will foster research that goes beyond fixing technologies to protect humans from harm, to also pursue enriching human experiences and connections through technology.
\end{abstract}



\maketitle

\section{Introduction}
\label{intro}

Much of social computing research today primarily focuses on addressing the flaws of current platforms, particularly in social media. However, this often limits exploring new, potentially more beneficial socio-technical systems. Indeed, today's research interventions are often confined within the bounds of platforms like Facebook and Instagram~\cite{facebook1, facebook2, facebook3, facebook4}, as opposed to pioneering new models, such as in research prototypes from two decades ago~\cite{miller2003movielens}. Furthermore, legislative approaches focus reactively on addressing problems on specific, existing platforms rather than proactively considering new possibilities ~\cite{press1, bill1, bill2, bill3} for the future of social technologies, furthering the view of social media as a fixed entity.

While prevention is important, focusing on positive and healthy potentials is equally important. We draw inspiration from positive psychology~\cite{seligman2000positive}, which arose from the understanding that people seek more than just relief from suffering; they aspire to live meaningful, fulfilling lives, nurturing their best qualities and enriching their experiences of love, work, and play ~\cite{seligman2019positive}. Decades of research in positive psychology have demonstrated that enhancing positive affect or life satisfaction is not merely about removing negative affect ~\cite{seligman2002authentic}.

Similar opportunities for positive impact are emerging in social computing research. To illustrate this, we consider social media as a representative example. Much research has understandably focused on addressing issues such as problematic social media use ~\cite{nesi2020impact, odgers2020annual, ito2020social, valkenburg2022advancing, van_der_wal_their_2022, sala2024social, davis2023supporting, tran2019modeling}, the adverse effects of social media on mental health ~\cite{choukas-bradley2023perfect, hamilton2022reexamining, jensen2019young, keles2020systematic, nesi2020impact, odgers2020annual, ito2020social, valkenburg2022advancing, sala2024social}, the effect of misinformation/disinformation on political dynamics ~\cite{suarez2021prevalence, wang2019systematic}, and the implications of privacy violations ~\cite{zhao2022understanding, hargittai2016can, Ali2022-cb, Agha2023-mu, WisniewskiPamela2022, razi2020privacy}. However, an additional, complementary approach allows us to more comprehensively address social media's core potential: fostering and amplifying human flourishing, or \textit{``living within an optimal range of human functioning, one that connotes
goodness, generativity, growth, and resilience,''} \cite{fredrickson2005positive} through the cultivation of positive emotions, engagement, relationships, meaning, purpose, and accomplishments, following the PERMA model of well-being ~\cite{seligman2011flourish}. To truly harness social media as technology centered on human flourishing, our research, design, and development must prioritize and be sensitive to both inter- and intra-personal growth and needs. We see great promise in integrating scientific knowledge and theories to make life more fulfilling into research on social technologies and in designing and creating new sociotechnical environments that better support human and societal flourishing through social interaction.

\subsection{Positive Social Technology (Positech)}
Given the historical evolution of social computing research and the rise of positive psychology, we propose \textbf{\textit{Positech}} as the theme of our workshop. Positech is a \textit{framework} that advocates for a shift in how we engage with social technology, emphasizing its potential to support and enhance human flourishing. It is characterized by focusing on the potential and opportunities for innovation and positive change that center on human flourishing rather than merely rectifying issues in humans or technology. Positech is both a framework and a set of efforts encompassing the development of technologies, legislative actions, industry initiatives, theoretical research, and meta-scientific research. 



Here, we lay out the distinct guiding questions of each adjacent field to Positech.
\begin{enumerate}[label=(\Alph*)]
    \item \textit{Positive Technology:} How can we use technology to increase emotional quality, engagement, and connectedness in personal experiences (e.g., therapy)?~\cite{riva2012positive, botella2012present, gaggioli2017positive}
    \item \textit{Positive Computing:} How can we account for well-being determinants such as autonomy, competence, relatedness, compassion, engagement, and meaning at every stage of interaction design?~\cite{calvo2014positive}
    \item \textit{Asset-based Approach:} How can we build upon users' strengths and self-determination in using and engaging with technology?~\cite{wong2020needs}
    \item \textit{Positive Social Technology (Positech):} How are we balancing the needs for fixing and innovating social technologies in supporting human flourishing in areas such as positive affect and meaning?
\end{enumerate}
An example outcome at the intersection of these fields could include research on a proof-of-concept for a new social media platform designed to explore new possibilities for increasing human connectedness. This platform would leverage people's intrinsic motivations to empathize with others to enhance the effectiveness of peer support.

Positech is distinct from positive technology or computing, which focuses on designing technology to improve personal well-being and quality of life. While principles of positive technology and computing guide systems creation, they represent only one aspect of Positech. Positech involves not just creating technology but also driving social technology development through broader ethical and societal dialogue about technology’s role. Unlike the asset-based approach, Positech does not define users' roles in technology use. It aims to maximize technology's potential in supporting universally valued aspects of human life, such as fostering connections and self-actualization, without making value judgments on individual capacities, which can vary across cultures and contexts. Positech also puts a particular emphasis on social technology.

It is important to note that within the Positech framework, we prioritize enriching interpersonal and personal experiences, viewing technology primarily as a supportive tool. While social technologies hold transformative potential for facilitating human connections, they should not replace them by default. Our goal is for technology to amplify direct human interactions and, where necessary, create new avenues for meaningful engagement. This ensures that technology \textit{enhances rather than substitutes for meaningful and authentic human connections}, maintaining our focus on supporting the core elements of human flourishing.

\subsection{Relevance to the CSCW Community}
We recognize that within the CSCW community, there are already numerous research initiatives that embody the principles of Positech, focusing on proactively enhancing human flourishing and well-being through technology ~\cite{to2023flourishing, wohn2018explaining, freeman2017social, maloney2020falling,freeman2021hugging,freeman2016revisiting,zytko2015enhancing,musick2021gaming,li2021interplay,li2023we,li2024beyond,liu2021significant,zhang2022auggie}. Our vision is not to present these ideas as new but to acknowledge and unify these scattered efforts under the Positech framework. By doing so, we aim to consolidate and amplify the impact of work that aligns with our goals:
\begin{enumerate}
    \item \textbf{Beyond \textit{fixing} people,} toward the goals of positive psychology such as human flourishing and well-being. This direction is in contrast to work that exclusively documents harms or seeks to effect behavior change without the intention of the user.
    \item \textbf{Beyond \textit{fixing} technology,} toward new paradigms and design patterns that prioritize positive psychology. This direction contrasts with work that aims to mitigate harms that exist on current platforms.
\end{enumerate}

These directions are not entirely separate but depict two different orientations toward our goals that best represent different types of research contributions. By focusing on both immediate and broader possibilities, we envision a future where CSCW is instrumental in unlocking the full potential of social technologies to contribute to human flourishing. However, it is crucial to acknowledge that social technology research has played a significant role in bringing us to our present understanding, and it is not the intention to diminish the value of these contributions. Rather, we call for a broader perspective that includes seeking new possibilities that would complement rather than replace current research momentum.
\section{Organizers}

\textbf{JaeWon Kim} is a PhD candidate at the University of Washington Information School. Her research focuses on understanding, designing, and building social technologies that center on meaningful social connections, especially for the youth.

\noindent\textbf{Lindsay Popowski} is a PhD candidate in the Stanford University Computer Science Department. Her research is in the field of social computing systems, with an emphasis on the design of social media. She works to design and build online spaces that facilitate unique social goals, such as strong interpersonal relationships, effort-sharing, and preserving conversational context.

\noindent\textbf{Anna Fang} is a PhD student at Carnegie Mellon University in the Human-Computer Interaction Institute, School of Computer Science. Her work is at the intersection of computational social science and mental well-being. She is predominantly interested in proactive approaches — those that are self-sustaining, self-correcting, or promote positive behaviors — for emotional and mental health, rather than retrospective handling of harm after it has occurred.

\noindent\textbf{Cassidy Pyle} is a Ph.D. Candidate at the University of Michigan School of Information. Her work examines how marginalized and stigmatized communities' interactions with and on socio-technical systems (e.g., social media, algorithms) may simultaneously afford \textit{and} constrain self-expression, emotional well-being, and access to college and career opportunities.

\noindent\textbf{Guo Freeman} is an Associate Professor of Human-Centered Computing at Clemson University. Her work focuses on how interactive technologies such as digital games, live streaming, social VR, and AI shape interpersonal relationships and group behavior; and how to design safe, inclusive, and supportive social VR spaces to mitigate emergent harassment risks.

\noindent\textbf{Ryan M. Kelly} is an Associate Professor in the School of Computing Technologies at RMIT University. His research focuses on the design and evaluation of communication technologies for fostering meaningful connections and alleviating social isolation. This includes designing for sensitive settings and vulnerable user groups, such as people with chronic health conditions or older adults living in long-term institutional care.

\noindent\textbf{Angela Y. Lee} is a Ph.D. Candidate in the Department of Communication at Stanford University. Her research investigates how communication technologies affect well-being. 

\noindent\textbf{Fannie Liu} is a VP Applied Research Lead on the Global Tech Applied Research AR/VR team at JPMorgan Chase \& Co. Her research involves the design of novel social experiences that promote social connection, collaboration, and well-being.

\noindent\textbf{Angela D. R. Smith} is an Assistant Professor in the School of Information at the University of Texas at Austin, where she co-leads the Research on Equity, Access, and incLusIon in Technology and societY (REALITY) Lab. Her research explores the design of socially responsible technology experiences by examining racial and cultural inequities.

\noindent\textbf{Alexandra To} is an Assistant Professor jointly appointed in the Art+Design department and the Khoury College of Computer Sciences at Northeastern University. She works in HCI, game design, critical race theory, and identity.

\noindent\textbf{Amy X. Zhang} is an Assistant Professor in the University of Washington Allen School of Computer Science and Engineering, where she leads the Social Futures Lab dedicated to reimagining social and collaborative systems to empower people and improve society. Her research areas broadly include social computing, CSCW, human-AI interaction, and HCI.
\section{Workshop Logistics and Agenda}

Our workshop at CSCW 2024 will bring together 15-30 researchers /practitioners/designers to discuss the area of Positech: aiming to envision new futures of positive social technology and strategize how to support each other's efforts. We aim to explore the overlapping fields of research that contribute to Positech, project a current landscape and future agenda, and plan how to build community and support our peers. More information on the agenda can be found on our workshop website: \href{https://positech-cscw-2024.github.io/}{https://positech-cscw-2024.github.io/}. We plan to publicize any outcome of our workshop.

\bibliographystyle{ACM-Reference-Format}
\balance
\bibliography{references}
\end{document}